\documentclass[a4paper,11pt]{article}
\usepackage{pos}
\usepackage{graphicx}   
\usepackage{subcaption} 

\title{Search for UHE neutrinos in the background of cosmic rays }

\author*[a]{Abha R. Khakurdikar}
\author[a,b]{Washington R. Carvalho Jr.}
\author[a]{Jörg R. Hörandel}

\affiliation[a]{IMAPP, Radboud University, Nijmegen, The Netherlands}

\affiliation[b]{Faculty of Physics, University of Warsaw, Warsaw, Poland}

\emailAdd{abha.khakurdikar@ru.nl}

\abstract{The main challenge in detecting ultra-high energy (UHE) neutrinos is discriminating a neutrino-induced shower in the background of showers initiated by ultra-high energy nuclei. The resulting shower development from neutrinos exhibits different characteristics from hadron-induced showers because neutrinos penetrate the atmosphere more deeply than hadrons.

This study focuses on simulations of highly inclined neutrino-induced showers above $75^\circ$ zenith angles, exploring an extensive energy range from $1 \text{EeV}$ to $120 \text{EeV}$. These simulated showers have different ranges of interaction depths corresponding to each zenith angle, presenting diverse detection challenges.

Our methodology utilises timing data from radio antennas for the shower front calculation for extensive air showers induced by neutrinos and nuclei. Furthermore, we incorporate signals obtained from Water Cherenkov detectors and the spatial distribution of stations registering signals in both Water Cherenkov detectors and radio antennas. We aim to classify neutrino-induced showers and background events stemming from nuclei by harnessing a decision tree classifier employing the Gini impurity method. Our framework yields excellent accuracy for separating the neutrinos from the background.

The findings of this study offer significant advancements in the domain of UHE neutrino detection, shedding light on astrophysical phenomena associated with these elusive particles amidst the complex background of UHE nuclei.}

\FullConference{%
  10th International Workshop on Acoustic and Radio EeV Neutrino Detection Activities - ARENA 2024\\
  11-14 June 2024\\
  Chicago, USA
}

\begin{document}
\maketitle

\section{Introduction}
The primary challenge in detecting ultra$-$high energy neutrinos is distinguishing a shower induced by neutrinos from the background of showers caused by UHE cosmic rays, which can be protons, heavy nuclei, and, to a much lesser extent, even photons. The simulations for downgoing neutrinos and hadrons$-$induced showers are produced at energies between $10^{18}$ and $10^{20}$ eV and at zenith angles between $75-85^\circ$. Neutrinos interact deep into the atmosphere. Hence, neutrino showers are generally much closer to the ground; hence, the shower wavefront is more curved than hadron$-$induced showers. We develop a method to reconstruct the shower fronts and fit the showerfront structure. Our methodology utilises timing data from radio antennas for the shower front calculation for extensive air showers induced by neutrinos and nuclei. The method is explained in Sec.\ref{showerfront}. Furthermore, we incorporate signals obtained from Water Cherenkov detectors and the spatial distribution of stations registering signals in both Water Cherenkov detectors and radio antennas. We aim to classify neutrino$-$induced showers and background events stemming from nuclei by harnessing a random forest classifier employing the Gini impurity method, explained in Sec.\ref{ml}. Our framework yields excellent accuracy for separating the neutrinos from the background. The findings of this study offer significant advancements in the domain of UHE neutrino detection, shedding light on astrophysical phenomena associated with these elusive particles amidst the complex background of UHE nuclei.

\section{Showerfront reconstruction for neutrinos}\label{showerfront}
Due to their weak interaction with matter, neutrinos can penetrate the Earth's atmosphere deep before producing extensive air showers. These neutrino-induced showers differ from those caused by hadronic cosmic rays as they occur closer to the ground, resulting in a more curved shower wavefront than the flatter wavefronts of hadron-induced showers. We reconstruct the shower front structure using the timing information from antenna signals. Geometrically, the arrival of air shower particles at lateral distances from the shower axis is delayed compared to a planar front. The time delay, calculated as $dtna = - (r_{\text{proj}} \sin(\theta_{\text{MC}}))/c$, where $r_{\text{proj}}$ is the projected distance to the shower axis, helps derive the corrected time signal for shower front reconstruction.

\begin{figure}[h!]
    \centering
    \begin{subfigure}{0.5\textwidth}
        \centering
        \includegraphics[width=\linewidth]{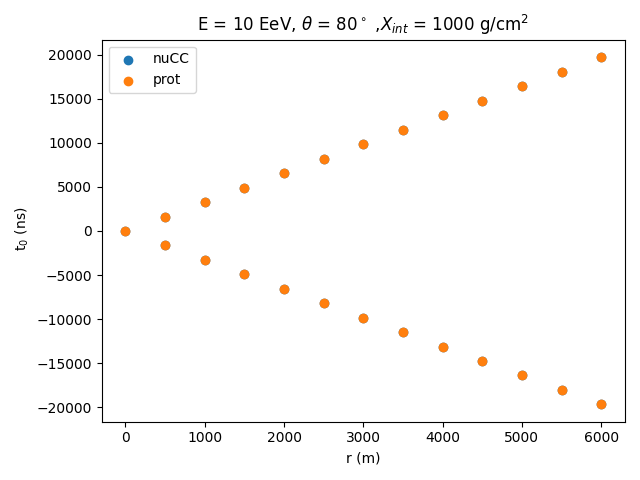}
        \caption{}
    \end{subfigure}%
    \hfill
    \begin{subfigure}{0.5\textwidth}
        \centering
        \includegraphics[width=\linewidth]{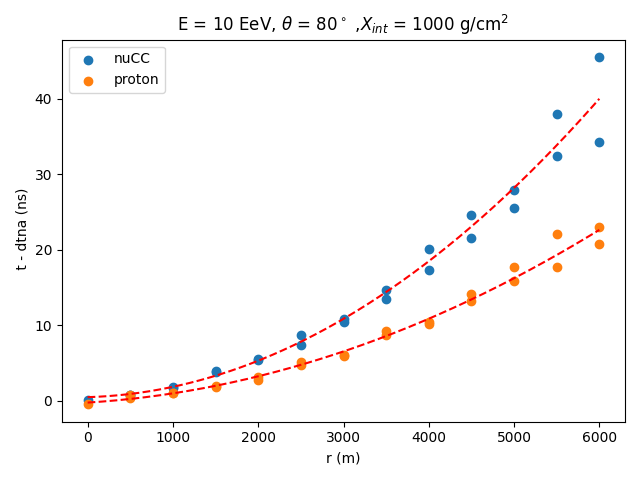}
        \caption{}
    \end{subfigure}
    \caption{(a) Signal time measured with respect to the distance of the antenna for proton and neutrino events. (b) Corrected time signal with respect to the distance of the antenna.}
    \label{fig:2dtime}
\end{figure}

Fig.\ref{fig:2dtime}(a) shows the initial signal time $t_o$ as a function of distance $r$ to the shower axis for proton and neutrino-induced showers, revealing no clear distinction between events. After correcting the time delay, Fig.\ref{fig:2dtime}(b) presents the corrected time as a function of $r$, showing a more curved structure for the neutrino event due to its deeper interaction depth of $1000 \text{ g/cm}^2$. Both the plots have an antenna spacing of $500 \text{ m}$. This greater curvature results from neutrino-induced showers starting deeper in the atmosphere than proton events, which develop higher up, leading to less curved wavefronts.

\begin{figure}[h!]
    \centering
    \includegraphics[width=0.7\textwidth]{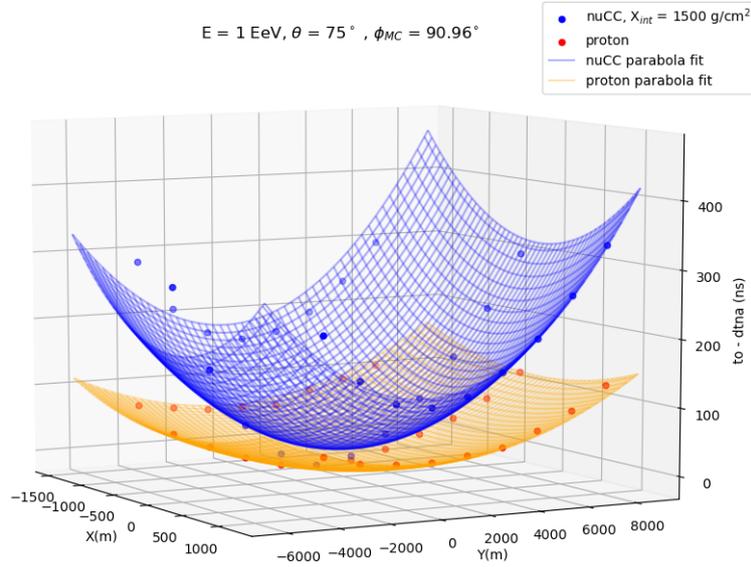}
    \caption{Corrected time signal with respect to the distance of the antenna position. The data points are fitted using the elliptical paraboloid equation.}
    \label{fig:paotime_fit}
\end{figure}

We also produced hadron and neutrino simulations with a $1.5 \text{km}$ spacing. The antenna response was unfolded in the simulations to reconstruct the electric field generated by the air shower. Partially cleaned noise and time jitter ($\sigma_t = 5\text{ ns}$) were added to the simulations. We calculated the time delay $dtna$ and corrected the time signal. The neutrino shower, developing deep into the atmosphere, exhibits a curved structure compared to the proton shower formed higher up. The fitting process involves modelling the relationship between the spatial coordinates $(x_i, y_i)$ and the corrected time $t_i = t_o - dtna$  using an elliptical paraboloid equation: $ax^2 + by^2 + cxy + dx + ey + f$. The goal is to determine the parameters $(a,b,c,d,e,f)$ that minimize the difference between the observed and predicted corrected time values. Fig.\ref{fig:paotime_fit} shows the fitted structure, where the neutrino shower (purple) with an interaction depth of $1500 \text{ g/cm}^2$ has a more curved front than the proton shower, despite both having the same primary energy and arrival direction. We obtained fit parameters for hadron (proton, helium, nitrogen, and iron) and neutrino simulations over the energy range of $1-120 \text{ EeV}$ and zenith angles of $75-85^\circ$. 

\section{Features for classification}\label{features_classification}

\begin{figure}[h!]
        \centering
        \includegraphics[width=0.7\textwidth]{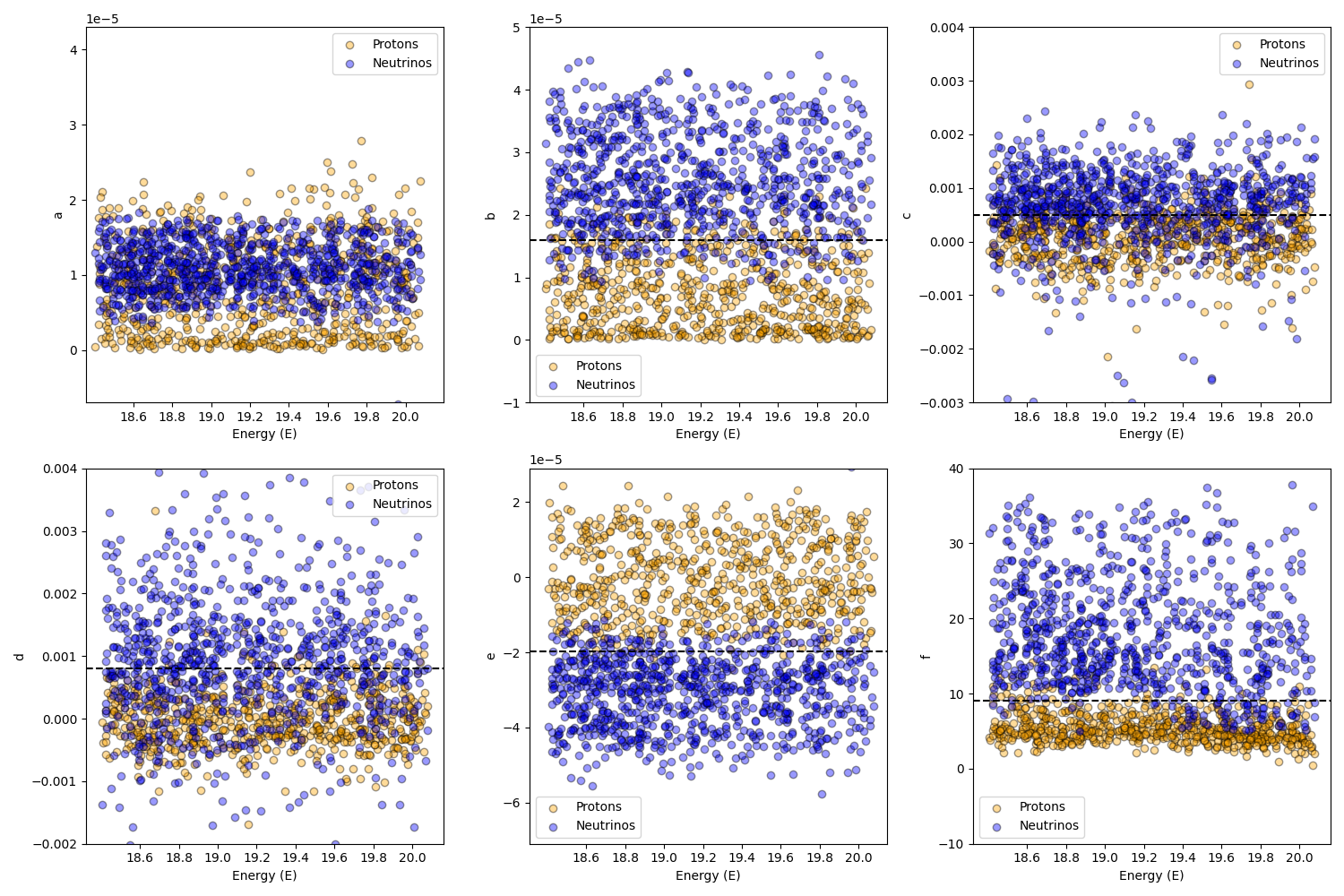}
        \caption{Fit parameters for the proton (orange) and neutrino (purple) events over the logarithmic energy range of $10^{18.4}-10^{20.1} \text{ eV}$.}
        \label{fig:energyfit}
    \end{figure}

The shower front reconstruction and fitting method are described in Sec.\ref{showerfront}. We obtain fit parameters for the hadron and neutrino events for the energy range of $1-120 \text{ EeV}$ and zenith angle range of $75-85^\circ$. Fig.\ref{fig:energyfit} shows the fit parameters $a,b,c$, from top left to right and $d,e,f$, from bottom left to right for proton and neutrino events over the energy range. We can see the separation in some parameters for neutrino and proton events.

\begin{figure}[h!]
        \centering
        \begin{subfigure}{0.48\textwidth}
            \centering
            \includegraphics[width=\linewidth]{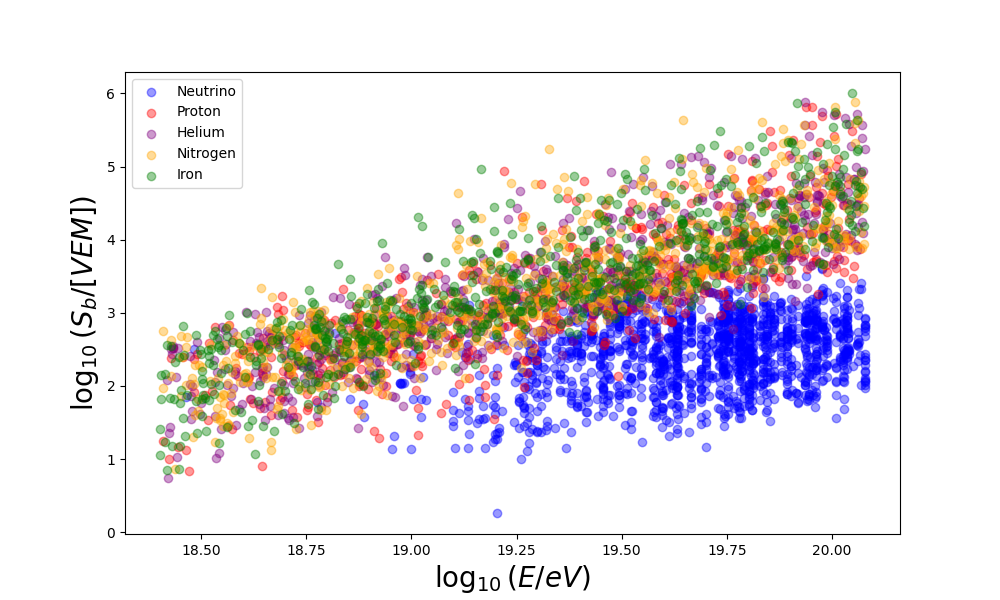}
            \caption{Total muon signal in an event over the energy range. Each data point represents one reconstructed event.}
            \label{fig:energySb}
         \end{subfigure}%
         \hfill
        \begin{subfigure}{0.48\textwidth}
            \centering
            \includegraphics[width=\linewidth]{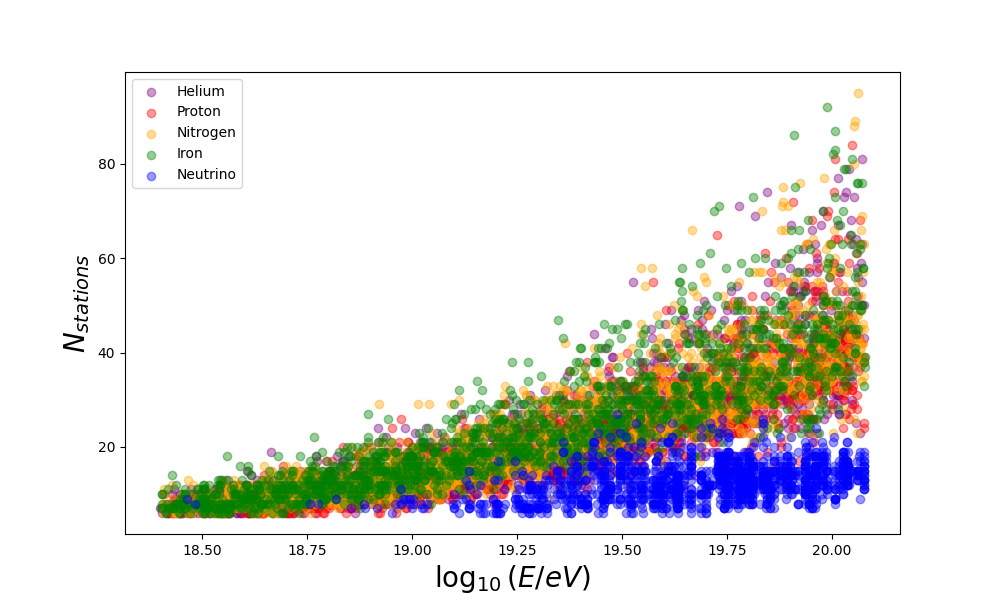}
            \caption{Number of stations with muon signal and radio signal over the energy range of $10^{18.4}-10^{20.1} \text{ eV}$.}
            \label{fig:energyNstat}
        \end{subfigure}
        \caption{Comparison of muon signal and number of stations over different energy ranges.}
    \end{figure}

 We also use the observable $S_b$, the total muon signal in an event. It is defined as,

    \begin{equation}
        S = \sum_{i} {S_{i}} \times \left( \frac{R_{i}}{R_\text{ref}} \right)^{b}
    \end{equation}

where the sum runs over the triggered stations, $S_i$ is the recorded signal in the $i-$th station at a distance $R_i$ from the reconstructed axis and $R_{\text{ref}}$ is a reference distance equal to $3500 \text{ m}$ for this analysis for the horizontal footprint of the inclined showers. The exponent $b$ is chosen equal to $4$ for maximizing the separation power between photons and hadrons. Fig.\ref{fig:energySb} illustrates the distribution of $S_b$ over the total energy range. The inclined hadron showers have significant muonic components at the detector level compared to neutrinos. Hence, the muon signal is higher for hadron showers.  The signal stations measured by muon signal and radio emission are used for the analysis. The events go through shower front reconstruction, and the events with the minimum number of $6$ signal stations with SNR $7.22$ are used for the classification. Fig.\ref{fig:energyNstat} shows the number of stations with muon signal and radio emission over the energy range $10^{18.4}-10^{20.1}$ eV.

\section{Machine Learning for the Classification of neutrinos and background}\label{ml}

\begin{figure}[h!]
    \centering
    \begin{subfigure}{0.48\textwidth}
        \centering
        \includegraphics[width=\linewidth]{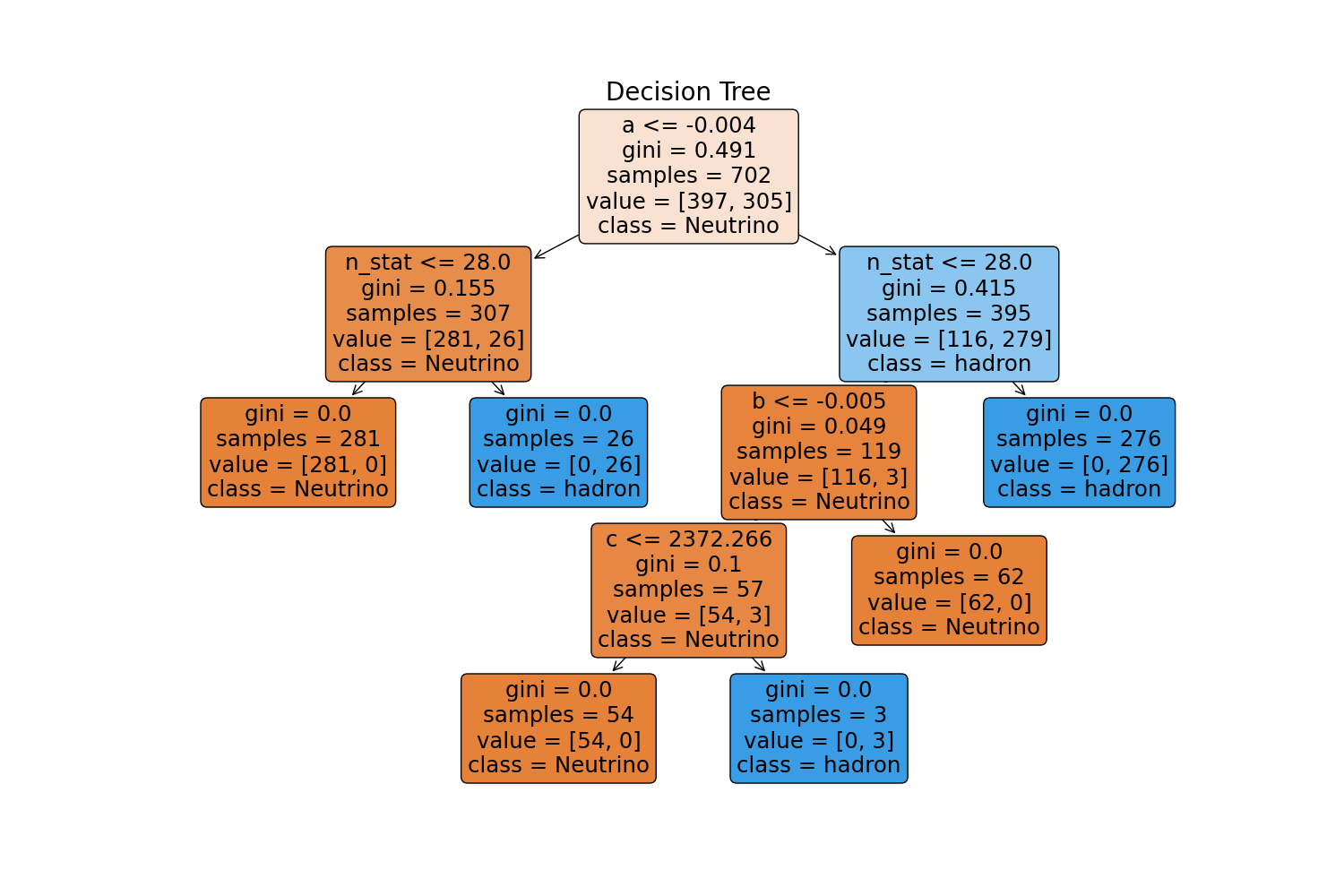}
        \caption{An example of a tree for tree depth 6 with Gini impurity$-$based data split. The data is divided into two classes, 'Neutrino' and 'Hadron', based on the feature threshold value and the Gini impurity.}
        \label{fig:treedepth}
    \end{subfigure}%
    \hfill
    \begin{subfigure}{0.48\textwidth}
        \centering
        \includegraphics[width=\linewidth]{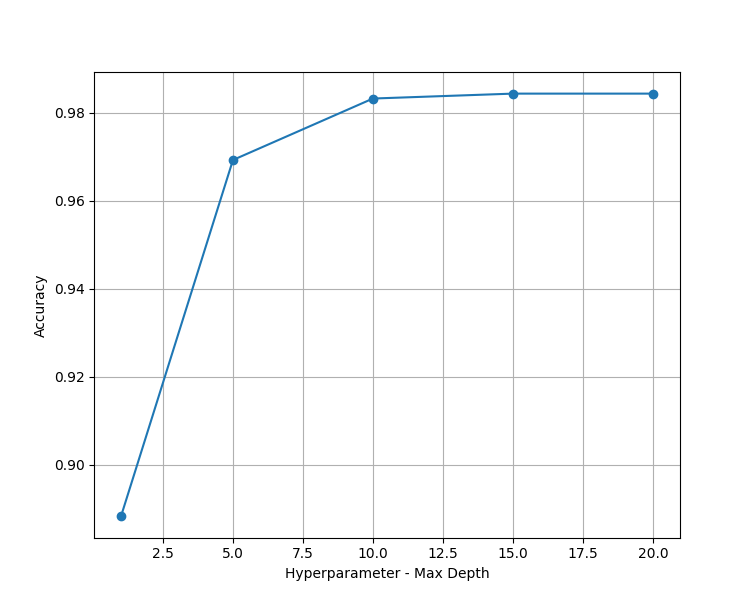}
        \caption{Model Accuracy as a function of maximum tree depth.}
        \label{fig:treedepth_accuracy}
    \end{subfigure}
    \caption{Comparison of tree structure and model accuracy as a function of tree depth in a binary classification task.}
\end{figure}
The dataset contains the primary energy, zenith angle, fit parameters, muon Signal, Number of stations and primary particle (neutrino or background (p, fe, he, n, photons)). After processing the data, the next step is to select the features and the target label for model training. The fit parameters( 'a', 'b', 'c', 'd', 'e', 'f'), Total muon Signal $S_b$, Number of stations $n_{\text{stat}}$ are chosen as the input features for the model. These features were chosen because they were relevant in distinguishing neutrinos and hadrons. The data is split into training and testing sets. Splitting the data into training and testing sets allows the model to be trained on one portion of the data (80\%) and tested on another portion (20\%).  

Hyperparameters are the parameters of a machine$-$learning model that need to be set before training begins. For Random Forest models, these hyperparameters define how the individual decision trees are built and how the ensemble of trees operates. We define a grid of hyperparameters for the Random Forest model, including options for the number of trees $(\text{n\_estimators})$, maximum depth $(\text{max\_depth})$, and minimum samples required to split a node $(\text{min\_samples\_split})$ and to be at a leaf node $(\text{min\_samples\_leaf})$. This cross$-$validation technique is used to find the best combination of hyperparameters by testing different combinations on the training data and selecting the one with the highest accuracy. The $\text{n\_estimators}$ hyperparameter specifies the number of decision trees in the ensemble. More trees generally improve performance because the model averages over more trees, reducing variance and improving stability. As more trees typically lead to better accuracy, increasing the number of trees makes the model more computationally expensive and time$-$consuming to train and predict. The numbers used in the initial grid search were $50, 100$ and $200.$

Decision trees evaluate split quality using metrics like Gini impurity and Entropy. Gini Impurity \cite{breiman1984classification} measures node impurity, aiming to minimize it for effective splits. The tree algorithm selects features and thresholds to reduce impurity, splitting nodes based on whether feature values meet certain criteria. A split with a high Gini gain is highly effective, making it desirable for tree construction. Entropy \cite{shannon1948mathematical}, derived from information theory, measures uncertainty and quantifies class label disorder. It calculates information gain to select the feature that reduces uncertainty the most. Although Gini impurity is computationally simpler, entropy can sometimes yield better splits due to its theoretical foundation.

The depth of a tree indicates its complexity and is defined as the maximum number of nodes from the root to the farthest leaf. As shown in Fig.\ref{fig:treedepth_accuracy}, accuracy often improves with increasing depth, but deeper trees are prone to overfitting. Hyperparameters like $max\_depth$, $min\_samples\_split$, and $min\_samples\_leaf$ control tree growth, affecting model generalization. Pruning, governed by the cost complexity criterion \( R_{\alpha}(T) = R(T) + \alpha \cdot |T| \), helps prevent overfitting by removing branches that offer minimal impurity reduction. After training, the model reveals the importance of each feature, which indicates how much each feature contributes to the classification. For the Gini method, feature importance is calculated by assessing how much each feature reduces impurity across all decision nodes. The resulting decision tree, built from these features and thresholds, provides decision rules that predict the $\texttt{primary\_particle}$ class labels for new instances.

\subsection{Model Evaluation}\label{modeleval}

\begin{figure}[h!]
    \centering
    \includegraphics[width=0.6\textwidth]{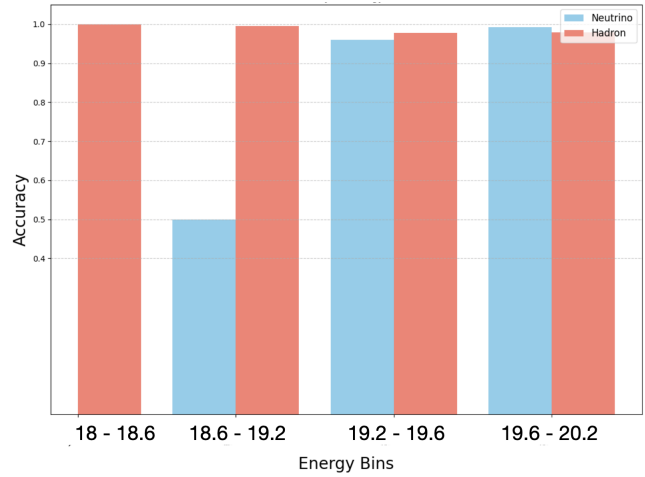}
    \caption{Model accuracy for different energy bins. Hadron events are classified with good accuracy across all energy ranges. Ultra-high energy neutrino events are also classified accurately. Neutrino events at low energy have low classification accuracy due to low statistics and the indistinguishable shower geometry from lower energy hadrons.}
    \label{fig:energybin}
\end{figure}

Model evaluation is essential in machine learning to assess how well a trained model performs on unseen data, ensuring its ability to generalize beyond the training set. The Random Forest model we developed was evaluated using multiple metrics, with accuracy being a simple yet widely used metric that measures overall correctness. The best model from the grid search was used to make predictions, achieving an accuracy of $0.9638$ for elliptical parabola fits in the ground plane. Different fitting equations were also tested, with parabola and hyperbola fits yielding lower accuracies of $0.861$ and $0.832$, respectively, as the shower plane geometry removes asymmetry structure, reducing classification accuracy. The accuracy was calculated as $\text{Accuracy} = \frac{\text{Number of Correct Predictions}}{\text{Total Number of Predictions}}$, and was optimized through an ensemble of $100$ trees and hyperparameter tuning. We further evaluated the model by splitting data into energy bins and calculating accuracy for each class (Neutrino and Hadron) separately. Lower accuracy for low-energy neutrinos was attributed to low statistics and the similar geometry of low-energy hadrons and high-interaction depth neutrino showers. 

\begin{figure}[h!]
    \centering
    \begin{subfigure}{0.48\textwidth}
        \centering
        \includegraphics[width=\linewidth]{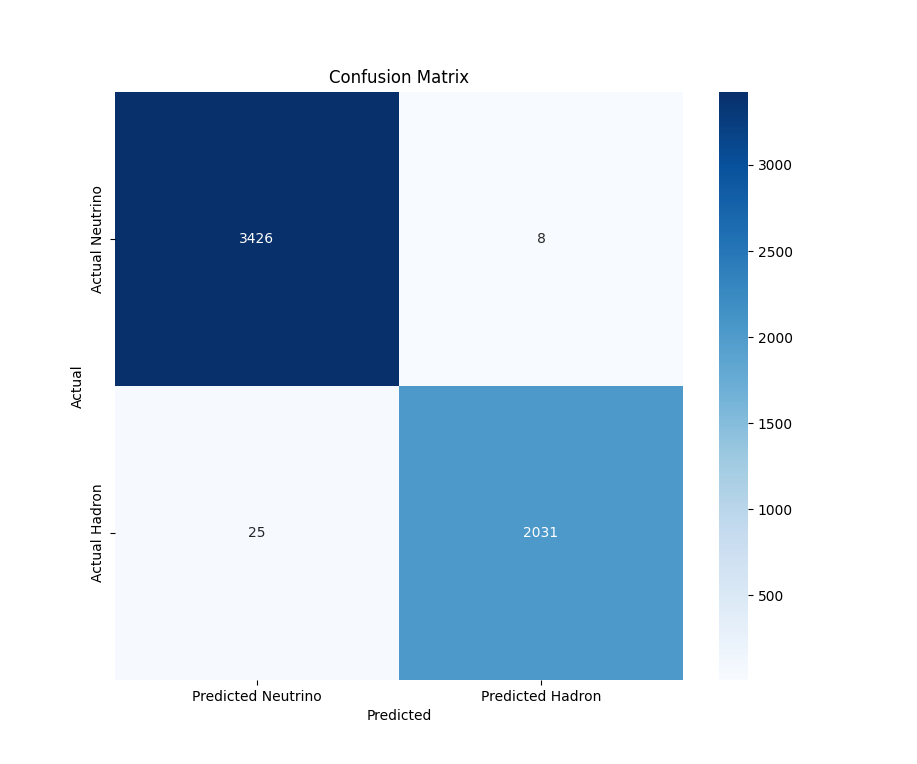}
        \caption{Confusion matrix for a binary classification for Neutrino and Hadron, showing the number of correct and incorrect predictions for each class.}
        \label{fig:confusionmat}
    \end{subfigure}%
    \hfill
    \begin{subfigure}{0.48\textwidth}
        \centering
        \includegraphics[width=\linewidth]{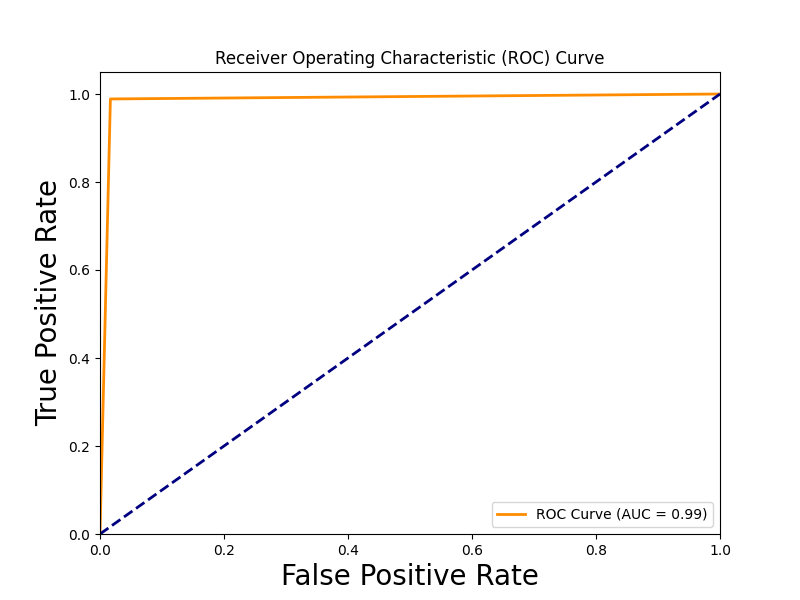}
        \caption{ROC Curve for the Random Forest Classifier performance for the binary classification of Neutrino and Hadron events. The AUC is $0.99$, indicating the classifier's good performance. The dotted line indicates the random classifier with an AUC of $0.5$.}
        \label{fig:roc}
    \end{subfigure}
    \caption{Comparison of classifier performance using confusion matrix and ROC curve for binary classification of Neutrino and Hadron events.}
\end{figure}

Fig.\ref{fig:energybin} shows model accuracy across energy bins, highlighting good classification accuracy for ultra-high energy neutrino events but challenges at lower energies. A confusion matrix (Fig.\ref{fig:confusionmat}) provides detailed insights into classification performance, showing an overall accuracy of $99.73\%$ for neutrino and proton classification. Additionally, ROC curves (Fig.\ref{fig:roc}) demonstrate the model's performance, with an AUC of $0.99$, indicating strong classification capability.

\begin{figure}[h!]
    \centering
    \begin{subfigure}{0.48\textwidth}
        \centering
        \includegraphics[width=\linewidth]{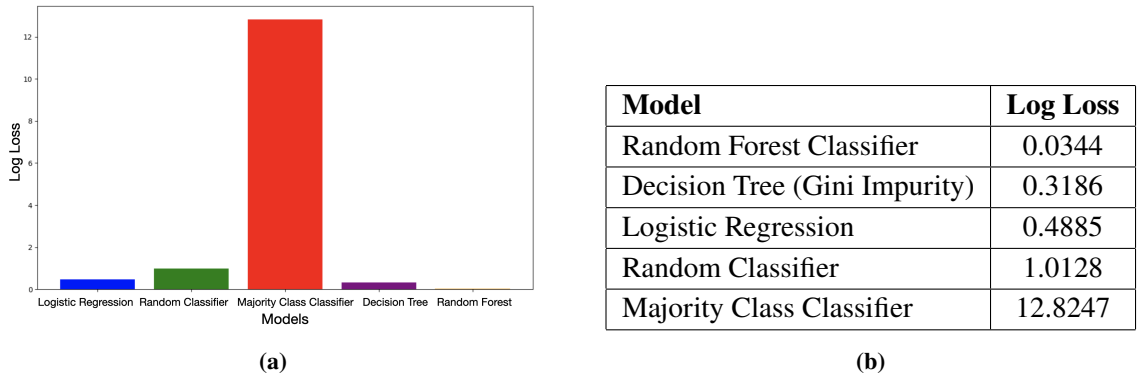}
        \caption{}
        \label{fig:logloss}
    \end{subfigure}%
    \hfill
    \begin{subfigure}{0.48\textwidth}
        \centering
        \begin{tabular}{|l|c|}
        \hline
        \textbf{Model} & \textbf{Log Loss} \\ \hline
        Random Forest Classifier & 0.0344 \\ \hline
        Decision Tree (Gini Impurity) & 0.3186 \\ \hline
        Logistic Regression & 0.4885 \\ \hline
        Random Classifier & 1.0128 \\ \hline
        Majority Class Classifier & 12.8247 \\ \hline
        \end{tabular}
        \caption{}
        \label{tab:logloss_comparison}
    \end{subfigure}
    \caption{(a) Log Loss calculations for different classifiers. The lower the Log Loss, the better the model's performance. (b) Log Loss values for different classifiers.}
\end{figure}
\begin{figure}[h!]
    \centering
    \begin{subfigure}{0.45\textwidth}
        \centering
        \includegraphics[width=\linewidth]{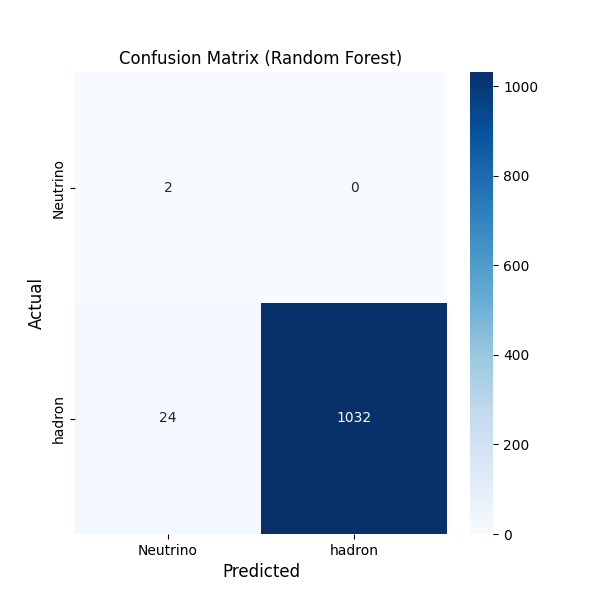}
        \caption{}
        \label{fig:2nuconf}
    \end{subfigure}%
    \hfill
    \begin{subfigure}{0.53\textwidth}
        \centering
        \includegraphics[width=\linewidth,height=6cm]{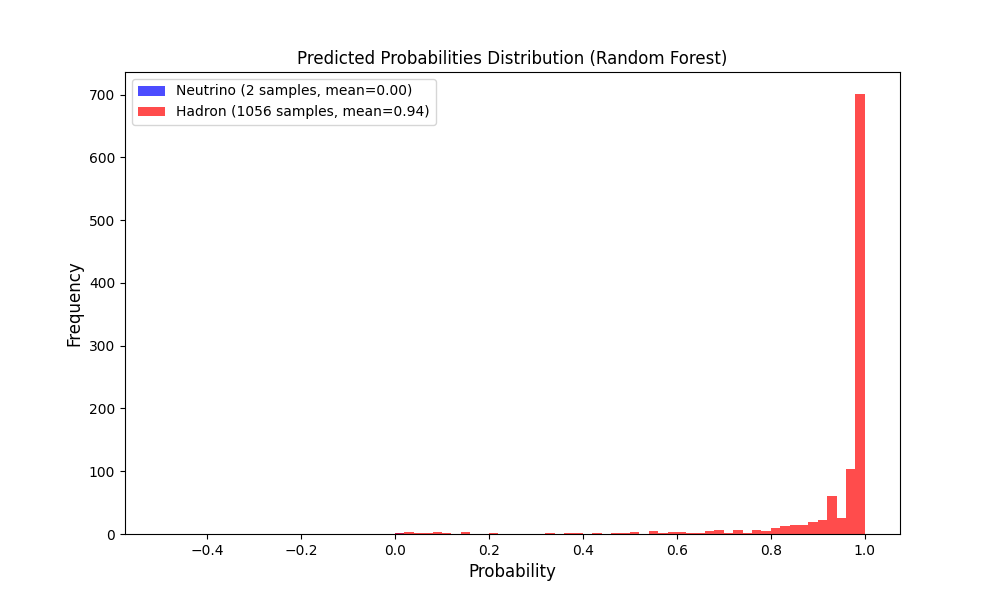}
        \caption{}
        \label{fig:probdist}
    \end{subfigure}
    \caption{(a) Actual and predicted hadron and neutrino events represented using Confusion matrix for the test set of $2$ neutrino and $1056$ hadron events. (b) Probability distribution of neutrino and hadron events. The class with the highest average probability is selected as the final prediction for that sample. $2$ neutrino events have an average probability of $0$, and the average probability for hadron events is $0.94$. }
\end{figure}

Log Loss (Fig.\ref{fig:logloss} and Table \ref{tab:logloss_comparison}) was also calculated, with the Random Forest achieving the lowest value of $0.03436$. In a further test using a set of $2$ neutrino and $1056$ hadron events, the model achieved $98.02\%$ accuracy, with a log loss of $0.1175$. Fig.\ref{fig:2nuconf} and Fig.\ref{fig:probdist} illustrate these events' confusion matrix and probability distribution, showing that the model effectively classified the rare neutrino events with high confidence.

\end{document}